\documentclass[aps,prb,twocolumn,floatfix,showpacs,superscriptaddress]{revtex4}
\pdfoutput=1

\usepackage{amsmath}
\usepackage{graphicx}  
\usepackage{color}
\begin{document}
\title{Mechanism for bipolar resistive switching in transition metal oxides}
\author{M.~J.~Rozenberg}
\affiliation{Laboratoire de Physique des Solides, UMR8502 
Universit\'e Paris-Sud, Orsay 91405, France}
\affiliation{Departamento de F\'{\i}sica J. J. Giambiagi, 
FCEN, Universidad de Buenos Aires,
Ciudad Universitaria Pab. I, 1428 Buenos Aires, Argentina}
\author{M.~J.~S\'anchez}
\affiliation{Centro At\'omico Bariloche and Instituto Balseiro, CNEA, 
8400 - San Carlos de Bariloche, Argentina}
\author{R.~Weht}
\affiliation{Gerencia de Investigaci\'on y Aplicaciones, CNEA, 1650 - San Mart\'{\i}n, 
Argentina}
\affiliation{Instituto Sabato, Universidad Nacional de San Mart\'{\i}n-CNEA,
Argentina}
\author{C.~Acha}
\affiliation{Departamento de F\'{\i}sica J. J. Giambiagi, 
FCEN, Universidad de Buenos Aires,
Ciudad Universitaria Pab. I, 1428 Buenos Aires, Argentina}
\author{F.~Gomez-Marlasca}
\affiliation{Gerencia de Investigaci\'on y Aplicaciones, CNEA, 1650 - San Mart\'{\i}n, 
Argentina}
\author{P.~Levy}
\affiliation{Gerencia de Investigaci\'on y Aplicaciones, CNEA, 1650 - San Mart\'{\i}n, 
Argentina}

\begin{abstract}
We introduce a model that accounts for the bipolar resistive 
switching phenomenom observed in transition metal oxides.
It qualitatively describes the electric field-enhanced migration of oxygen vacancies at the nano-scale. 
The numerical study of the model predicts that strong electric fields develop in the
highly resistive dielectric-electrode interfaces, leading to a spatially inhomogeneous
oxygen vacancies distribution and a concomitant resistive switching effect. 
The theoretical results qualitatively 
reproduce non-trivial resistance hysteresis experiments that we also report, 
providing key validation to our model.
\end{abstract}
\maketitle

\section{Introduction}

There is a great deal of experimental activity currently devoted
to explore new technologies for the next generation of electronic
memory devices~\cite{meijer}.
Among various promising options, the resistive random access memory (RRAM), which is
based on the resistive switching (RS) phenomenon, has emerged as a
preeminent candidate~\cite{waser,sawa}. 
The RS effect is
a large, reversible and nonvolatile change in the resistance after 
the application of voltage or current pulses.
The typical RRAM system has a capacitor-like structure composed of 
insulating or semiconducting materials sandwiched between two metal electrodes. 
RS has been observed in a wide variety of systems, such as
simple and complex oxides, organic compounds, etc~\cite{janousch,sawa}. 
However, there are specific characteristics of the switching behavior 
observed in each type of material.
In the case of binary oxides, which are highly insulating it is believed
that the RS effect may be due to the formation and rupture of 
conductive filaments within the insulating media~\cite{szot,inoue,fujiwara}. 
In contrast, in the more conducting or semiconducting complex oxides with perovskite structures,
such as doped cuprates and manganites, the relevance of oxygen vacancies is often
invoked. 
Despite a bursting body of experimental data that is rapidly becoming
available~\cite{baek,liu,beck,watanabe,choi} the precise mechanism 
behind the physical effect of RS remains elusive.
A few qualitative models have been proposed emphasizing different 
aspects: electric field-induced defect migration~\cite{szot, baikalov, nian}, phase
separation~\cite{tulina}, tunneling across interfacial domains~\cite{ris},
control of Shottky barrier's height~\cite{sawa,jeong2}, etc.
A general consensus has emerged on the empirical relevance of 
three key features: (i) a highly spatially inhomogeneous conduction in 
the low resistive state, (ii) the existence of a significant number of 
oxygen vacancy defects 
and (iii) a preeminent role played by the interfaces, namely, 
the regions of the oxide that are near each of the metallic electrodes
which often form Schottky barriers.
Here, we shall introduce and study the behavior of a simple model that 
incorporates, at a qualitative level, those three features. 
By means of a numerical simulation we shall show that the 
model
correctly reproduces key non trivial hysteresis cycles observed 
in experiments on perovskite-type transition metal
oxides (TMO).

\section{Model}

Several experiments revealed that the
conduction in the low resistive state is highly inhomogeneous and dominated by one dimensional paths 
that are associated with enhanced conduction channels~\cite{szot,fujiwara}. 
These paths would be created upon an 
initial application
of strong electric fields, that bring the dielectric
close to its breakdown point.
Thus, we shall assume that the electric
transport is dominated by a single conductive path embedded within a more insulating host. 

\begin{figure}
\centerline{\includegraphics[width=6.5cm]{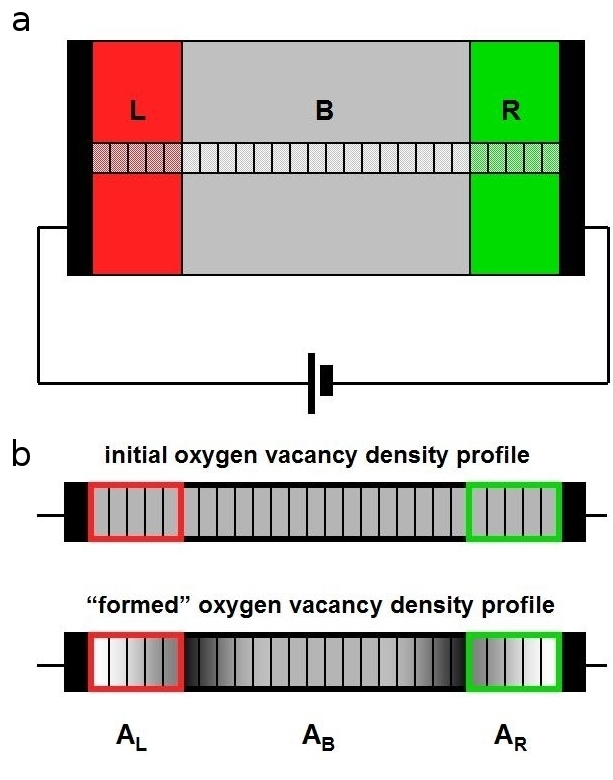}} 
\caption{a) Schematic model with a single conductive channel within the dielectric.
The three regions $L$, $R$ and $B$ correspond to the two high resistance interfaces 
and the more conductive central bulk, respectively. The small boxes indicate the domains. 
b) Detailed scheme of the conductive path. The grayscale
qualitatively depicts the variation in the oxygen vacancy concentration through the channel 
(darker corresponds to higher concentration). The top figure shows the initial state with
uniformly distributed oxygen vacancies, and the bottom one shows the inhomogeneous distribution
after the first few ``forming'' voltage cycles (see text).
}
\label{fig1}
\end{figure}

The second important feature incorporated into our model is the 
relevance of defects within the dielectric. Several experiments point to
a preeminent role played by oxygen vacancies \cite{szot,seong,nian,tsui}.
Moreover, it is a universal and salient feature of TMO
that their resistivity is dramatically affected by the precise oxygen stoichiometry. 
One may thus expect 
that oxygen vacancy concentration may be the most significant parameter
controlling the {\it local} resistivity, $\rho$, of a given material. 
This feature is included in our model by assuming that each nano-domain
of the path is characterized by a certain concentration of oxygen vacancy defects, $\delta$.
We adopt the most simple linear relation $\rho \propto \delta$,
which follows from the fact that
in  TMO perovskites the presence of oxygen vacancies 
severely disrupts the electronic conduction properties.
Nevertheless, we emphasize that the specific form of $\rho(\delta)$
is not crucial for the results that we shall describe later on. 

The third important feature that our model incorporates is 
the key role played by the interfaces \cite{ris}, namely, the regions of the
dielectric that are in physical proximity to the metallic electrodes. 
There is growing evidence that these are the regions where the RS takes place 
\cite{baikalov,chen,quintero,jeong2,sawa2,fujii}. 

Our model is schematically shown in Fig.~\ref{fig1} and consists of a single
conductive channel within a more insulating dielectric, which is represented by 
a one dimensional resistive network on $N$ links. The first and last $N_I$ links correspond
to high resistance interfacial regions next to the external electrode, 
and the central $N - 2 N_I$ links describe the bulk section.
Each link is characterized by a certain concentration of oxygen vacancies, 
which determines the resistivity of the link. 
They may be physically associated to small domains of nanoscopic dimensions
which may actually correspond to grains  of the polycrystalline oxide. We take
$\rho_i = A_\alpha \delta_i $, with $\alpha = B$ if $i$ is in the bulk ($N_I < i \leq N-N_I$), 
$\alpha = L$ if $i$ is in the left interface ($i \leq N_I$)
and $\alpha = R$ if  $i$ is in the right interface ($N-N_I < i \leq N$).
In our study we set $N=100$ and $N_I=10$.
The following equation specifies how the vacancies diffuse through the 
network domains under an external voltage,
\begin{equation}
p_{ab} = \delta_a (1-\delta_b) \exp(-V_0 + \Delta V_{a}) \;.
\label{proba}
\end{equation}
It gives the probability for transfer of 
vacancies from domain $a$ to a nearest neighbor domain $b$.
The probability is proportional to the concentration of vacancies
present in domain $a$ and to the concentration of ``available vacancy
sites'' at the target domain. 
The Arrhenius factor $\exp(-V_0)$, is controlled by a 
dimensionless constant, $V_0$, related to the activation energy for vacancies diffusion. 
The important
factor $\exp(\Delta V_{a})$ models the enhancement (or suppression)
of the diffusive process due to the local electric field at domain $a$.

From Eq.(\ref{proba}),
a constant $V_0$
leads to an initially constant distribution
$\delta_i=\delta_o$ for all $i$. 
The value of the 
initial constant concentration, $\delta_o$
must be much smaller than one, since it physically represents the concentration
of defects (oxygen vacancies) within a domain. We adopt $\delta_o = 10^{-4}$.

Similarly to actual resistive switching experiments, we simulate the applied
voltage protocol $V(t)$ by linear ramps that follow the sequence 
$0 \to +V_{\rm max}\to 0 \to -V_{\rm max} \to 0$ (our convention
is that the right electrode is grounded). The duration is of $s$ time steps. 
The sequence may be repeated a number of cycles $n$, for a total 
duration $\tau_{\rm max} = n s$. In our simulations we choose $V_{0}/V_{\rm max}=0.016$,
that provides a non-negligible but slow diffusive contribution 
to the evolution of $\delta_i$ with respect to the total time
duration of the simulations (ie, the total number of time steps). We set
$V_{\rm max}=1000$ that provides for a sufficiently large electric stress.
Our qualitative results are rather robust with respect to the choice of
model parameters, a detailed systematic study of their dependence is left
for future work.
 
The coefficients $A_B$, $A_R$ and $A_L$ still remain to be specified.
With no loss of generality, we fix the value of the bulk coefficient to unity $A_B=1$ 
and leave the interfacial $A_R$ and $A_L$ free. In this initial study we shall concentrate
in the symmetric case, $A_R = A_L$, that corresponds to most common experimental devices.

The numerical simulations are performed through the
following steps: (i) at each simulation time step $t$ ($1 \leq t \leq \tau_{\rm max}$)  
a given external voltage $V(t)$ is applied to the resistive network 
(the electrodes are assumed perfect conductors). 
The current through the system is 
computed as $I(t) = V(t) / R_T$, with $R_T$ the total (ie, two terminal) 
resistance 
\begin{equation}
R_T = c \sum_{i=1}^{N} {\rho_i} = \sum_\alpha \sum_{i\in \alpha}{A_\alpha \delta_i},
\label{resistance}
\end{equation} 
where $\alpha = R,~ B,~ L$ denotes the three network regions, and $c$
denotes an unessential geometrical constant related to the dimensions of
the domains (which we set to unity). 
(ii) we compute the local voltage profile $V_i(t) = I(t) \rho_i$ and the voltages drops 
${\Delta V}_i (t) = V_{i+1}(t) - V_i(t)$.
(iii) we use Eq.~(\ref{proba}) to compute all the oxygen vacancy transfers between nearest neighboring 
domains, and update the values $\delta_i(t)$ to a new set of concentrations $\delta_i(t+1)$.
(iv) we use these new values to recompute the current at $\tau_{t+1}$ under the applied voltage
$V(t+1)$ as indicated in the first step. 
\begin{figure}
\centerline{\includegraphics[width=7.5cm]{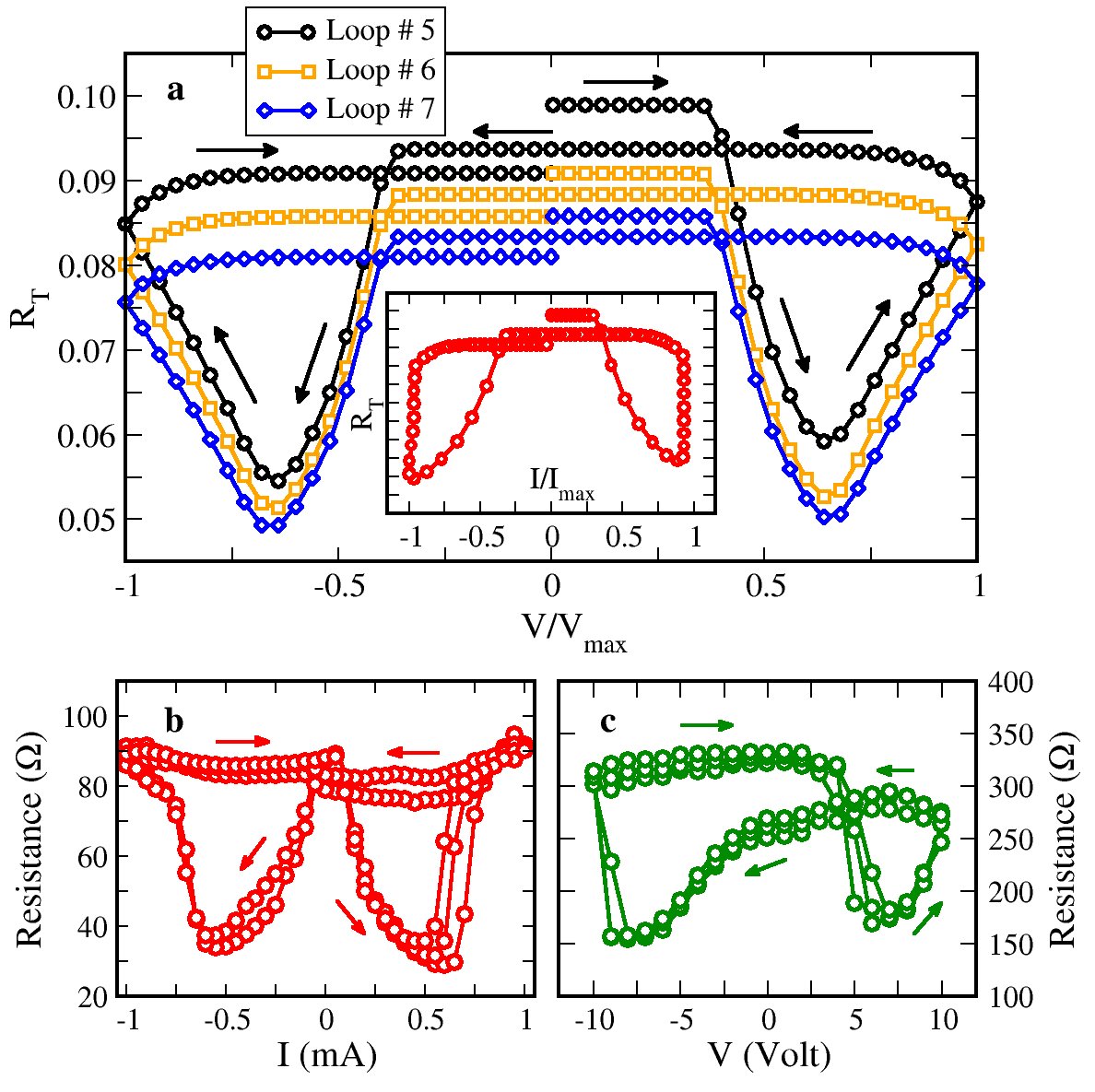}} 
\caption{Top panel: Resistive hysteresis loop (R$_T$ vs V) for the subsequent voltage cycles number 
5 , 6 and 7.
({\bf a}). Inset: the R$_T$ vs I hysteresis (loop number 5)
shows qualitatively similar results (the vertical axis scale is the same as in the main figure). 
Bottom panel: experimental hysteresis loops measured
in manganite ({\bf b}) and cuprate ({\bf c}) devices.
The first one was pulsed in current control mode while the second in
voltage mode. The details of the experimental procedures are 
described in Refs. \onlinecite{quintero} and \onlinecite{acha}, respectively.}
\label{fig2}
\end{figure}

\section{Numerical Results and Comparison to Experiments}

We now turn to the discussion of our results. We set
$A_R = A_L = 1000 >> A_B =1$, which is consistent with
our experimental data and with previous reports in bulk
and thin films of conducting perovskites~\cite{nian,quintero,acha}.
In Fig.~\ref{fig2} we show the results for the hysteresis loop of the total resistance $R_T$.
The different data curves in Fig.~\ref{fig2}a display the results in
subsequent voltage cycles 5, 6, and 7. 
The inset of Fig.~\ref{fig2}a, show that the hysteresis as a function of current remains
qualitatively similar. 
We note that during the first few initial cycles the resistance shows 
non-repetitive memory effects that converge to a hysteresis loop with a stable shape. 
Interestingly, this is reminiscent of the initial ``forming'' that
experimental samples seem to require in order to start displaying reproducible 
switching effects.
The peculiar type of hysteresis loop that we obtain has been already reported by the
Houston group \cite{ignatiev} in experiments on (Pr,Ca)MnO$_3$  manganite systems, 
where it has been termed ``table with legs''. It is evidently a non-trivial effect and 
we have experimentally reproduced it in both,
a related manganite (Pr,La,Ca)MnO$_3$
and a cuprate (YBa$_2$Cu$_3$O$_{7-x}$) 
sample, as shown in the bottom panels of Fig.~\ref{fig2}. 

There are several features worth pointing out:
During each voltage protocol  loop, there is a clear variation of the resistance $R_T$
between  a rather broad maximum for a large range of $V$ (ie, the ``table''), and two 
relatively narrow minima (ie, the ``legs''). These maximum and minima correspond to the
high and low resistance states, $R_T^{HI}$ and $R_T^{LO}$.
The $R_T(V)$ loops are approximately symmetric in $V$ which reflects the left-right
symmetry of the system. 
Throughout the voltage loop, the system begins 
in the initial $R_T^{HI}$ state and 
undergoes the sequence of resistance changes
$\to R_T^{LO+} \to R_T^{HI+}$ under positive bias, and then
$\to R_T^{LO-} \to R_T^{HI-}$ under negative bias. The final state, at zero bias 
is $R_T^{HI}$, very close but not identical to the previous initial state.
Interestingly, a similar small drift is also observed in the experimental data.

The qualitative agreement between our prediction of the ``table
with legs'' and the experimentally observed hysteresis loops provides a significant validation
for our model. Therefore, to gain physical insight into the mechanism of the RS effect we shall discuss in detail 
the evolution of  the vacancies distribution
under the applied voltage protocol.
\begin{figure}
\centerline{\includegraphics[width=7cm]{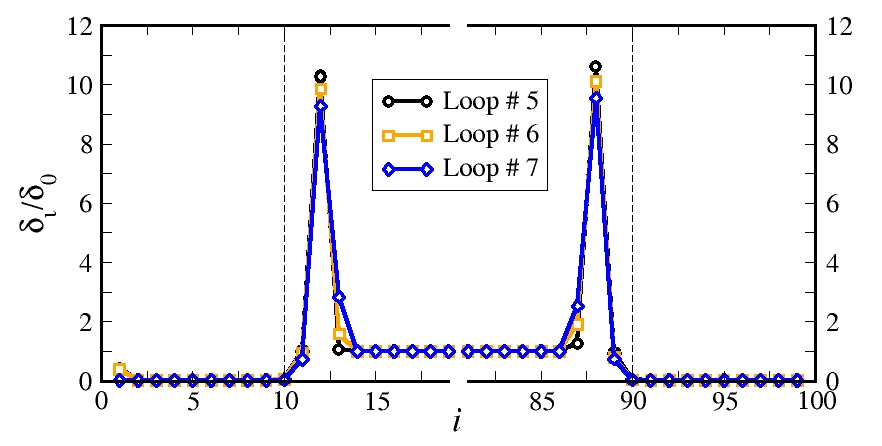}} 
\caption{Density concentration profiles normalized to the initial uniform density value 
$\delta_i / \delta_o$, at the beginning of voltage
cycles number 5, 6 and 7 in the symmetric system.}
\label{fig3}
\end{figure}
In Fig.~\ref{fig3} 
we show successive snapshots of 
the oxygen vacancy concentration profile $\delta_i$ at the beginning of loops
5, 6 and 7. 
Recalling that the initial equilibrium distribution of oxygen vacancy concentration 
is uniform $\delta_i = \delta_o$, these curves reveal that,
under the action of the repetitive voltage cycling, the $\delta_i$ evolve towards a 
new stable distribution. The salient features of the profile $\delta_i$ are a 
significant depletion in the interfacial regions and a strong accumulation peaks at
both internal boundaries between the bulk and the interfacial regions.
The reason can be understood as follows. The largest
electric fields occur at the two interfacial regions since initially their resistance 
is much larger than the bulk one ($A_L$, $A_R >>  A_B$). 
Therefore, oxygen vacancy migration is enhanced
in the interfacial regions, with the ions moving either towards the electrodes  or the
bulk, depending on the direction of the applied voltage.
When the ions reach the metallic electrodes they start to pile-up \cite{nian}. 
On the other hand, the vacancies that migrate towards
the bulk eventually leave the interfacial region and enter the bulk.
There, their diffusion virtually stops, since the 
electric fields in the more conducting bulk are much smaller. 
Successive initial cycles yield a cumulative effect, with a depletion of the 
interfacial regions (and some pile-up at the edges) and a concomitant 
accumulation at the bulk side of the interfacial/bulk boundary.
Significantly this accumulation, as seen in Fig.~\ref{fig3}, is quite large and narrow leading to a substantial 
increase in the local resistance. This feature will be a key to understand the
origin of the legs of the hysteresis loop, which we shall consider next.

\begin{figure}
\centerline{\includegraphics[width=8cm]{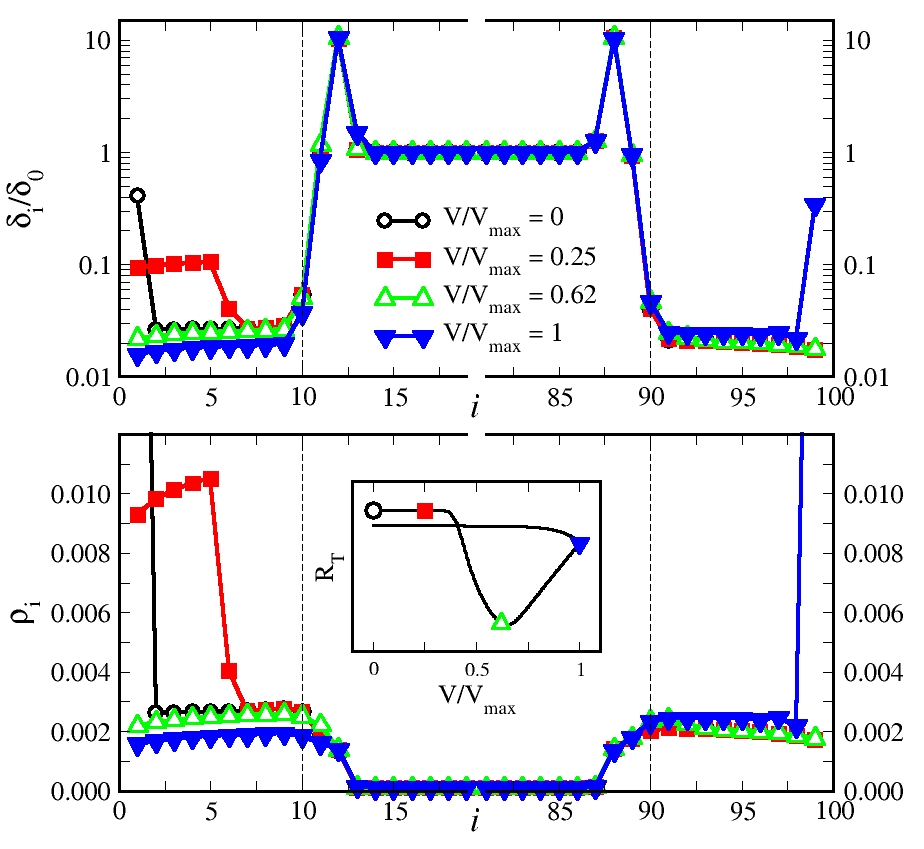}} 
\caption{Top (semi-log scale): Snapshots of the density concentration profiles for the symmetric configuration,
normalized to the initial uniform density value 
$\delta_i / \delta_o$, during the hysteresis cycle (first half). 
Bottom: The corresponding profiles of local resistance $\rho_i$ for the same snapshots.
Inset: Position of the snapshots in the (first half) hysteresis loop. 
}
\label{fig4}
\end{figure}
In Fig.~\ref{fig4} we show snapshots, during the first half of the voltage cycle,
of the vacancy concentration and local resistance profiles
in the interfacial regions and their boundaries with the bulk. These are the
active regions of the system where the electric-field-enhanced migration takes place.
Notice that since the current $I(t)$ is uniform along the conductive path, the
local electric fields are directly proportional to the local resistance.
Let's start at $V$=0, from the state at the beginning of a cycle.
Interestingly, the $\rho_i$ profile indicates significant 
electric fields in both interfacial
regions which extends into the neighboring bulk. Thus the depletion of vacancies in the interfaces 
and the accumulation at the boundaries are such that 
compensate for the difference between the respective
$A$ coefficients, to yield similar electric fields across the boundaries of the regions. 
Yet, at the very left end of the system there is a small pile-up of vacancies  which, as soon as the voltage is ramped up, 
will translate into the largest local fields and initiate the ionic migration.
At $V/V_{\rm max}$=0.25, we observe a snapshot of the migration of the vacancies across the left interfacial region 
towards the bulk. From the resistance expression Eq.~(\ref{resistance}), so long the vacancies remain within the
interfacial region, the system remains in $R_T^{HI}$ state.
At $V/V_{\rm max}$=0.62, the vacancies have moved out the left interfacial region and entered the bulk, where their
migration suddenly stops. Once in the bulk, the contribution to the total resistance of these 
migrating vacancies is reduced ($A_B << A_L$), thus the system reaches the $R_T^{LO}$ state (leg of the table).
Notice that in this state, the largest fields occur at the boundaries of the bulk and the interfacial regions.
Thus, as $V$ is further increased, the left interfacial region depletes further and therefore the
voltage drop gets lower there. In contrast, on the right side the electric fields are further enhanced
and there is now a migration of vacancies from the accumulation peak of the bulk towards the right
interfacial region. This leads to an increase of the total resistance and, at the maximal voltage
$V/V_{\rm max}$=1, we find that the vacancies have entered the interfacial region and already piled-up at the
right end. Thus, the system is back to $R_T^{HI}$.
The voltage protocol continues with the decrease of $V$ back to zero but keeping the same (positive)
polarity. Therefore no significant change in the $\delta_i$ and $\rho_i$ profiles occurs, and
half of the table with legs is already formed. When the negative
polarity part of the cycle begins a similar analysis follows, since the distribution of vacancy
concentrations is a mirror image of the initial one. This forms the other half of the table.

\begin{figure*}
\centerline{\includegraphics[width=18cm]{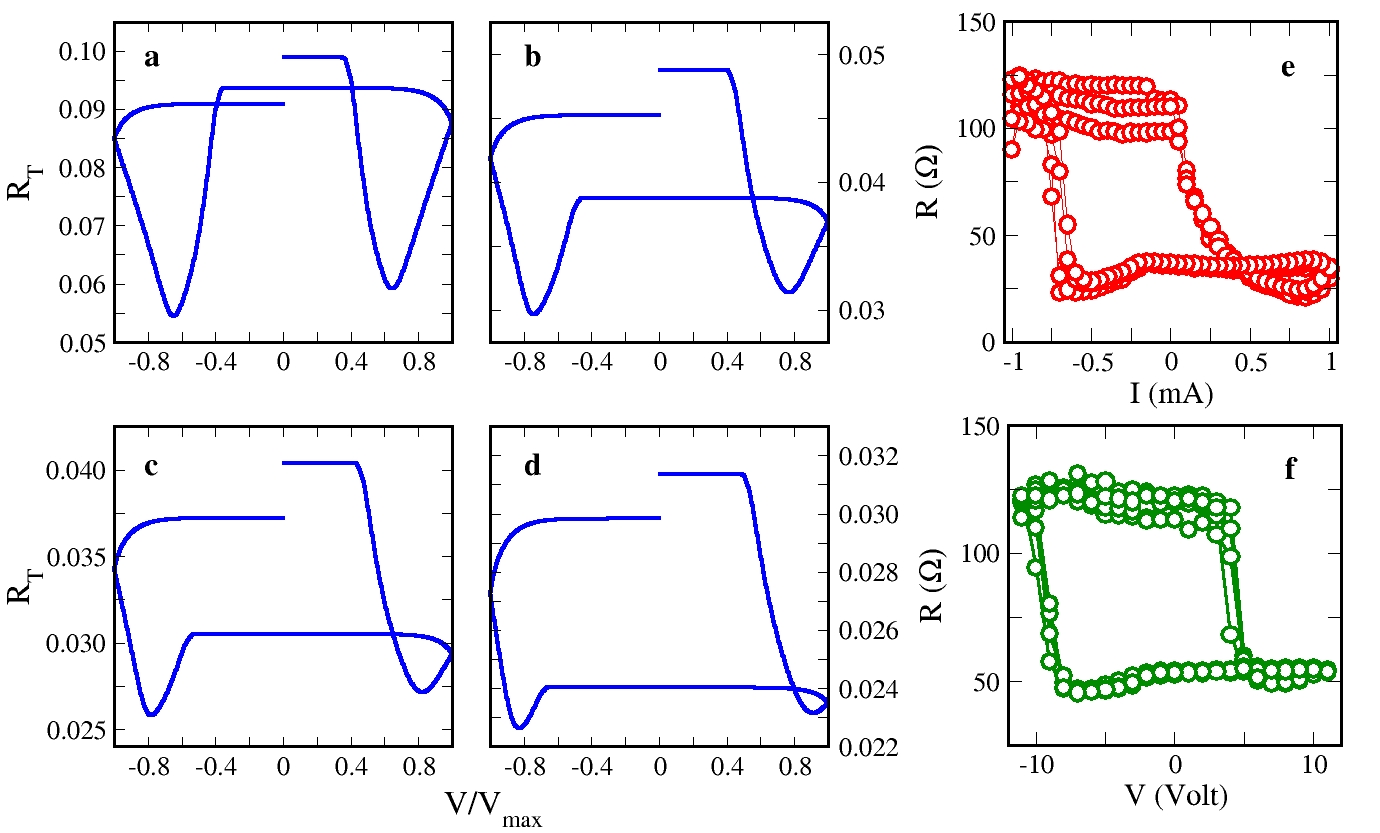}}
\caption{Resistive hysteresis loops obtained for an increasing degree of
asymmetry. $A_L$ = 1000 and $A_R$ = 1000, 100, 50 and 25 in panels 
{\bf a}, {\bf b}, {\bf c} and {\bf d} respectively. 
The right panels show experimental data for a manganite ({\bf e})
and a cuprate sample ({\bf f}) that were rendered asymetric by means of intense and fixed polarity pulsing.
The former was pulsed in current control mode and the latter in
voltage control, similarly as in Refs.~\onlinecite{quintero}
and \onlinecite{acha}, respectively.
 Notice the qualitative similarity of data of panels {\bf e} and 
{\bf f} with panel {\bf d}; and of data in Fig. 2b with panel {\bf a}.
The data of Fig. 2c seem to be in between the results of
panel {\bf a} and {\bf b}.
}
\label{fig5}
\end{figure*}

To complete our study we consider the effect of introducing
an asymmetry in the model parameters $A_R$ and $A_L$.
In Fig.~\ref{fig5} we display the results for the hysteresis
loops upon increasing the asymmetry. The results of the
simulations show the gradual evolution of the ``table with
legs'' towards a conventional rectangular hysteresis cycle~\cite{ignatiev}.
Asymmetry in experimental samples can either be due
to the fabrication process, as for instance, in the obvious
case of using different type of metal for the electrodes.
Though our samples were fabricated in a symmetric
configuration, we induced a significant asymmetry
by vigorous pulsing with a given polarity (ie, poling)
before measuring the resistance hysteresis characteristic.
Significantly, as shown in Fig.~\ref{fig5}, we find that the
experimental data obtained in an asymmetric manganite
sample and the cuprate, induced by intensive same polarity pulsing,
are in good qualitative agreement with our simulations.
These results provides additional validation to our model.

\section{Conclusions}

To conclude, our results put on solid theoretical grounds the key role 
played by oxygen vacancies in the mechanism of resistive switching in TMO.
They also provide valuable insights, predicting a non-trivial spatial profile of the oxygen 
vacancy distribution which may be of help for device design.
An exciting idea for future work is to explore the possibility of using
atomistic, first principles, calculations to study properties of electrode - transition metal oxide interfaces 
to estimate the parameters of the model and provide guidance in the material choice
for actual memory devices. 

\section*{ACKNOWLEDGMENTS}

Support from CONICET (grants PIP 5254/05 and PIP 112-200801-00047) and ANCTyP 
(grants PICT 483/06 and PICT 837/07) is gratefully acknowledged.

\end{document}